\documentclass[prb,twocolumn,showpacs,superscriptaddress,preprintnumbers,amsmath,amssymb]{revtex4-1}
\usepackage{graphicx} \usepackage{dcolumn} \usepackage{bm} \usepackage{amsmath} \usepackage{color}
\usepackage{hyperref} \usepackage[utf8]{inputenc} \usepackage[T1]{fontenc}

\begin{document}


\title{Softened magnetic excitations in the $s=3/2$ distorted triangular antiferromagnet
$\alpha$-CaCr$_2$O$_4$}

\author{Dirk Wulferding} \affiliation{Institute for Condensed Matter
Physics, Technical University of Braunschweig, Mendelssohnstr. 3, D-38106 Braunschweig, Germany}

\author{Kwang-Yong Choi} \affiliation{Department of Physics, Chung-Ang University, 221
Huksuk-Dong, Dongjak-Gu, Seoul 156-756, Republic of Korea}
\affiliation{Institute for Condensed Matter
Physics, Technical University of Braunschweig, Mendelssohnstr. 3, D-38106 Braunschweig, Germany}

\author{Peter Lemmens} \affiliation{Institute for Condensed Matter
Physics, Technical University of Braunschweig, Mendelssohnstr. 3, D-38106 Braunschweig, Germany}

\author{Alexey N. Ponomaryov} \affiliation{Department of Physics, Chung-Ang University, 221
Huksuk-Dong, Dongjak-Gu, Seoul 156-756, Republic of Korea}

\author{Johan van Tol} \affiliation{Department of Chemistry and Biochemistry, Florida State University, Tallahassee, Florida 32306, USA}
\affiliation{National High Magnetic Field Laboratory, Florida State University, Tallahassee, Florida 32310, USA}

\author{A. T. M. Nazmul Islam}
\affiliation{Helmholtz Zentrum Berlin f\"{u}r Materialien und Energie, Hahn-Meitner-Platz 1, D-14109 Berlin, Germany}

\author{Sandor Toth}
\affiliation{Helmholtz Zentrum Berlin f\"{u}r Materialien und Energie, Hahn-Meitner-Platz 1, D-14109 Berlin, Germany}
\affiliation{Institut f\"{u}r Festk\"{o}rperphysik, Technische Universit\"{a}t Berlin, Hardenbergstr. 36, D-10623 Berlin, Germany}

\author{Bella Lake}
\affiliation{Helmholtz Zentrum Berlin f\"{u}r Materialien und Energie, Hahn-Meitner-Platz 1, D-14109 Berlin, Germany}
\affiliation{Institut f\"{u}r Festk\"{o}rperphysik, Technische Universit\"{a}t Berlin, Hardenbergstr. 36, D-10623 Berlin, Germany}

\date{\today}

\begin{abstract}
The spin dynamics and magnetic excitations of the slightly distorted triangular $s=3/2$ system $\alpha$-CaCr$_2$O$_4$ are investigated by means of Raman spectroscopy and electron spin resonance (ESR) to elucidate its peculiar magnetic properties. Two-magnon excitations in circular $RL$ symmetry show a multi-maximum structure with a dominant spectral weight at low energies.
The temperature dependence of the ESR linewidth is described by a critical broadening $\Delta H_{pp}(T)\propto (T-T_N)^{-p}$ with the exponent $p=0.38(5)-0.48(3)$ for temperatures  above $T_N = 42.6$ K. The exponent is much smaller than that of other $s=3/2$ triangular lattices. This is ascribed to soft roton-like modes, indicative of the instability of a helical 120$^\circ$ phase.
As an origin we discuss a complex spin topology formed by four inequivalent nearest neighbor and sizable next-nearest neighbor interactions.

\end{abstract}


\maketitle

\section{Introduction}

The possibility of observing exotic states of matter with unusual excitations has created a vast interest in systems with competing interactions. Promising candidates are geometrically frustrated quantum spin systems, such as two-dimensional (2D) kagome or triangular lattices.

Very early on, P. W. Anderson proposed a spin liquid ground state for the $s=1/2$ triangular lattice antiferromagnet, a state in which magnetic order is suppressed and the spins remain fluctuating down to lowest temperature.~\cite{anderson} However, it is now established that a nearest-neighbor Heisenberg antiferromagnet on a perfect triangular lattice forms a non-collinear 120$^\circ$ ordered ground state.~\cite{zheng,Bernu}
Compounds that realize a triangular lattice generally show small deviations from the perfect models studied theoretically. On the one hand, the existence of two slightly different nearest neighbor exchange interactions $J_{nn}$ due to the formation of isosceles triangles can drive the system into a state of long range order, spin liquid, gapped dimer, or valence bond crystal, depending on the ratio of $J_{nn1}$ and $J_{nn2}$.~\cite{doretto} On the other hand, an inclusion of next-nearest neighbor interactions $J_{nnn}$ can induce a transition to a collinear phase when $J_{nnn}/J_{nn}> 0.125$.~\cite{Wheeler} This raises important questions about the ground state and its low energy excitations if these two mentioned factors are integrated in a single lattice.

The distorted delafossite compound $\alpha$-CaCr$_2$O$_4$ is ideally suited to address such issues because the Cr$^{3+}$ ($s=3/2$) ions form a slightly distorted triangular lattice with substantial
next-nearest neighbor interactions. There are four inequivalent nearest neighbor Cr$^{3+}$-Cr$^{3+}$ distances varying between 2.889~\AA  and 2.939~\AA.~\cite{pausch,toth} This leads to a unique spin network consisting of two straight and two zig-zag chains, see Ref.~\cite{toth} and the inset in the lower panel of Fig. 1. Measurements of the magnetic susceptibility yield the mean strength of the exchange interaction $J_{mean} = 6.48$ meV ($\mathrel{\widehat{=}} 75$ K), the Curie-Weiss temperature $\Theta_{CW} = -564$ K, and the effective magnetic moment $\mu_{eff} = 3.68 \mu_B$.~\cite{toth} An inelastic neutron scattering (INS) study details the magnetic parameters:
the average nearest neighbor exchange interaction $J_{nn}^{av} =8.8(8)$~meV and the average next-nearest neighbor exchange interaction $J_{nnn}^{av}=0.69$~meV.~\cite{toth-prl} Below $T_N=42.6$ K, the compound orders magnetically with a helical 120$^\circ$ spin structure and the ordering wave vector $\vec{k} = (0, 0.3317, 0)$, i.e., close to the commensurate $\vec{k} = (0, 1/3, 0)$.~\cite{chapon} This seems to prove the robustness of the 120$^\circ$ phase despite lattice distortions and higher-order interactions. However, INS experiments further unveiled the softening of low energy modes with roton-like minima at wavevectors different from $\vec{k} = (0, 1/3, 0)$.~\cite{toth-prl,perkins}
This was taken as evidence for an instability in $\alpha$-CaCr$_2$O$_4$ due to its vicinity to a new magnetic phase, possibly described by multiple spin waves.

In proximity to a phase boundary, enhanced quantum fluctuations and thereby order by disorder can lead to a whole spectrum of effects for frustrated magnets. To fully explore these aspects, it is necessary to investigate magnetic excitations by employing spectroscopic methods complementary to neutron scattering. Raman spectroscopy is a well-suited experimental choice because it allows to directly probe the two-magnon spectrum and to investigate the effect of these interactions on the spectral weight. In particular, the theoretical study on the dynamics of magnons in noncollinear antiferromagnets~\cite{perkins,chernyshev} suggests a substantial amount of magnon-magnon interactions. In addition, electron spin resonance (ESR) provides valuable information on the evolution of short-range spin correlations.

In this paper, we report on softened magnetic excitations with a multi-maximum structure and unusual spin dynamics in $\alpha$-CaCr$_2$O$_4$, which are not expected for a simple helical 120$^\circ$ triangular lattice. This suggests that the magnetic structure is largely described by a helical 120$^\circ$ ordering but low-energy excitations cannot be understood within the framework of a simple 2D triangular antiferromagnet. This is ascribed to a complex spin topology resulting from the combined effects of inequivalent nearest-neighbor and next-nearest neighbor interactions.

\section{Experimental details}

Single crystals of $\alpha$-CaCr$_2$O$_4$ were grown by the optical floating zone method.~\cite{islam-unpub} The crystals were cut into thin plates along the crystallographic $a$ axis. Thereby, optically flat $bc$ planes were obtained with average surface dimensions of 5 $\times$ 2 mm$^2$. Raman scattering experiments were performed in quasi-backscattering geometry, i.e. the $\vec{k}$ vector of the incident and scattered light perpendicular to the sample surface. A $\lambda = 532$ nm solid state laser was used with the laser power set to 7.5 mW with a spot diameter approximately 100 $\mu$m to avoid heating effects. All measurements were carried out in an evacuated closed cycle cryostat in the temperature range from 10 K to 295 K. The spectra were collected via a triple spectrometer (Dilor-XY-500) by a liquid nitrogen cooled CCD (Horiba Jobin-Yvon, Spectrum One CCD-3000V).

High-frequency ESR experiments were performed at 240 GHz using the quasi-optical spectrometer that has been developed at the National High Magnetic Field Laboratory with a sweepable 12~T superconducting magnet.
Our spectrometer employs a superheterodyne detection scheme with high-frequency Schottky diode mixers and a lock-in amplifier for field modulation. Thus, the field derivative of a microwave absorption signal can be recorded as a function of an external magnetic field.~\cite{Hans}

\section{Results and Discussion}

\begin{figure}
\label{figure1}
\centering
\includegraphics[width=8cm]{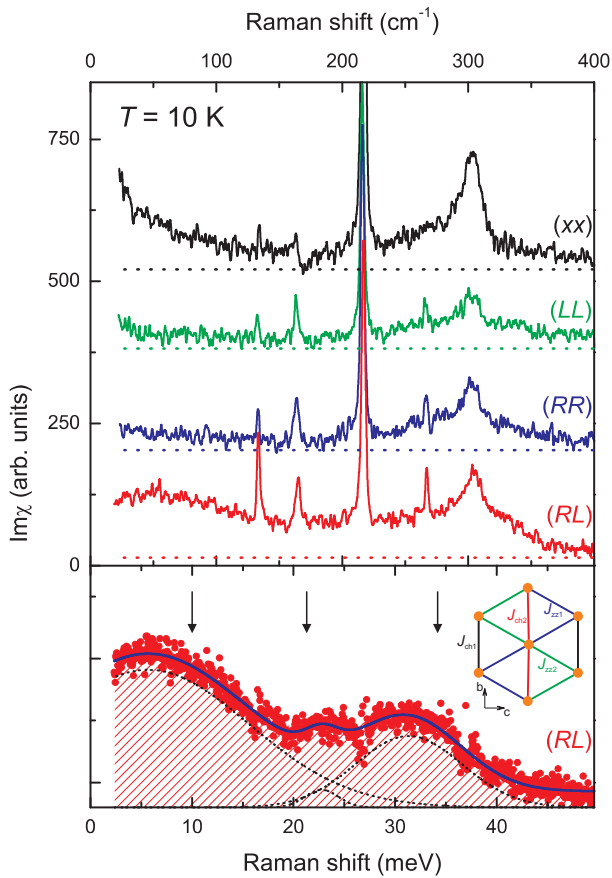}
\caption{(Colour online) Bose corrected Raman spectra of $\alpha$-CaCr$_2$O$_4$ obtained for different light polarizations $xx$, $LL$, $RR$, and $RL$ at $T=10$ K. The spectra are shifted in intensity for clarity. The dashed horizontal lines mark the background for each spectrum. The lower panel displays the spectrum obtained in $RL$ polarization with the phonon lines subtracted together with fit curves of three gaussian lines (dashed black lines) and a sum of the fits (solid blue line). The arrows mark twice the roton energies obtained from INS.~\cite{toth-prl} The inset shows the distorted triangular arrangement of the Cr$^{3+}$ ions and their exchange geometry.}
\end{figure}

The crystal structure of $\alpha$-CaCr$_2$O$_4$ is orthorhombic (space group $Pmmn$) with four formula units per primitive cell.~\cite{pausch} The factor group analysis yields for this crystal symmetry 36 Raman-active phonon modes: $\Gamma_{Raman} = 11 \cdot A_g + 6 \cdot B_{1g} + 11 \cdot B_{2g} + 8 \cdot B_{3g}$. The corresponding Raman tensors are given by:

\begin{center}

\begin{widetext} \mbox{A$_{g}$=$\begin{pmatrix} a & 0 & 0\\ 0 & b & 0\\ 0 & 0 & c\\
\end{pmatrix}$

, B$_{1g}$=$\begin{pmatrix} 0 & d & 0\\ d & 0 & 0\\ 0 & 0 & 0\\ \end{pmatrix}$

, B$_{2g}$=$\begin{pmatrix} 0 & 0 & e\\ 0 & 0 & 0\\ e & 0 & 0\\ \end{pmatrix}$

, B$_{3g}$=$\begin{pmatrix} 0 & 0 & 0\\ 0 & 0 & f\\ 0 & f & 0\\ \end{pmatrix}$.} \end{widetext}
\end{center}

All Raman scattering experiments are performed within the $bc$ plane of the crystal due to the sample's thin, plate-like geometry. This only allows the observation of A$_g$ and B$_{3g}$ symmetry components in the quasi-backscattering geometry.
Single crystals of $\alpha$-CaCr$_2$O$_4$ host three crystallographic twins with their unit cells rotated by an angle of $\pm$ 60$^\circ$ around the $a$ axis. Therefore, neither the $b$ nor the $c$ axis is fixed with respect to the laboratory reference of frame and it is not possible to distinguish between A$_g$ and B$_{3g}$ phonon modes. We can clearly observe 13 out of the expected 19 Raman-active phonon modes. The discrepancy can be due to possible overlapping with phonons of larger intensities as well as a lack of pronounced phonon intensity.

Hereafter, we will focus on magnetic Raman scattering. In contrast to square lattice antiferromagnets, the magnetic scattering from frustrated, undistorted triangular lattices should be independent of the in-plane scattering geometry of the incoming and outgoing light.~\cite{perkins} In the case of $\alpha$-CaCr$_2$O$_4$, the small distortion might induce a minute anisotropy in the excitation spectra for different scattering geometries. However, the twinning will ultimately lead to an averaging of these anisotropies.

In the upper panel of Figure 1, we show the low energy range (20 -- 400 cm$^{-1}$) of Raman spectra obtained in different polarizations at $T = 10$ K. We observe a rather weakly structured background for the spectra in $LL$ and $RR$ polarization and an appreciable background scattering with a quasi-elastic tail in $xx$ polarization. In $RL$ polarization, an enhanced background is visible. For clarity, base lines are added to each curve. Subtracting all phonon modes from the $RL$ spectrum yields the spectral weight shown in the lower panel of Figure 1. As the background is very broad and mostly pronounced in the $RL$ polarization and shows no dependence on laser wavelength (not shown here), we exclude a phonon density of states as a possible origin. Recent INS studies on powder samples revealed magnetic excitations in $\alpha$-CaCr$_2$O$_4$ ranging from lowest energies up to around 40 meV ($\approx 320$ cm$^{-1}$).~\cite{toth-prl} In particular, the shift of the spectral weight toward lower energies is comparable to the monotonic increase of the momentum averaged neutron scattering intensity. Judging from the energy range and spectral form (compare the lower panel of Fig.~1 to Fig.~2(a) in Ref.~\cite{toth-prl}), we ascribe this signal to a two-magnon like scattering.

As mentioned above, the two-magnon scattering of an ideal triangular lattice shows no directional dependence. This selection rule can be checked by inspecting the circular polarized spectra. The $RR$ and $LL$ polarizations correspond to $xx-yy-i(xy+yx)$ and $xx-yy+i(xy+yx)$, respectively, with $R = x - iy$ and $L = x + iy$. Since the magnetic scattering intensity is the same in $xx$ and $yy$ polarization, we expect little to no magnetic contribution in both $RR$ and $LL$ polarization. Consistently, the magnetic background is hardly discernible. In contrast, the $RL$ polarization corresponds to $xx+yy+i(xy-yx)$ and thus the magnetic contribution from the different parallel polarizations add up. Indeed, we observe a pronounced magnetic continuum in the $RL$ scattering configuration.

For the case of the ideal triangular lattice, the $xx$ and the $RL$ polarization spectra should scale with each other. Overall, both the $xx$ and $RL$ polarization spectra show a similar shift of the spectral weight toward lower energies. However, the $RL$ polarization spectrum possesses several maxima around 5.5, 23, and 32~meV while the $xx$ polarization spectrum shows a strong quasi-elastic like tail for energies below 20~meV with a broad maximum around 33~meV. The latter is explained by the huge spectral weight at $\omega\approx 0$ observed by INS.~\cite{toth-prl} In addition, the more pronounced structure in the $RL$ spectrum is due to deviations from the ideal triangular lattice and implies a complex exchange topology. We note that similar observations of enhanced scattering intensity of helical states in $RL$ polarizations were reported in the case of the topological insulator Bi$_2$Se$_3$ although the underlying scattering mechanisms are different from each other.~\cite{gnezdilov}

\begin{figure}
\label{figure2}
\centering
\includegraphics[width=8cm]{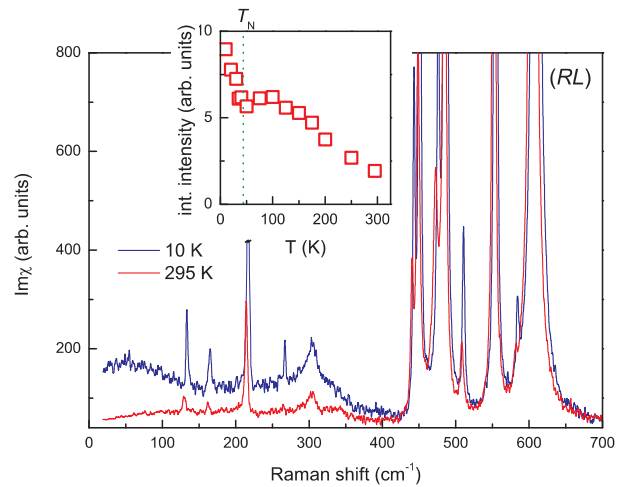}
\caption{(Colour online) Comparison of the Bose corrected spectra obtained in $RL$ polarization at 10 K (blue curve) and 295 K (red curve). The inset shows the integrated intensity of the spectral weight as shown in the lower panel of Fig. 1 over temperature.}
\end{figure}

Figure 2 compares two spectra obtained at $T = 10$ K and 295 K in $RL$ polarization. The phonon modes show a moderate hardening together with a gain in intensity upon cooling. Overall, the temperature dependence of the phonon parameters can be explained by lattice anharmonicities. The magnetic continuum, observed strongest in $RL$ symmetry, increases in spectral weight with decreasing temperature. For a square lattice, with increasing temperature through $T_N$ the spectrum dampens and the peak energy renormalizes strongly.~\cite{Choi} However, there is no substantial change in the width and the position of the two-magnon continuum in a wide temperature range for our case. In contrast, only a gradual suppression of the spectral weight is associated with the characteristic feature of a frustrated system, that is, the persistence of locally correlated magnetic states. The inset in Fig.~2 plots the integrated intensity of the magnetic continuum over temperature. Around $T_N=43$ K, a sudden change in slope occurs, marking the crossover from short range magnetic correlations to long range magnetic order.

The two-magnon scattering mechanism for conventional square lattices is well established. The bare two-magnon density of states has a van-Hove singularity due to a flat magnetic dispersion at the Brillouin zone boundary. Magnon-magnon interactions broaden the singularity and shift the maximum intensity down to $E_{max} = J(2zs-1)$. Here, $J$ is the magnetic exchange constant, $z$ is the number of magnetic bonds (i.e. nearest neighbors) and $s$ is the spin quantum number. This picture is well confirmed by experimental results.~\cite{Choi} However, the same argument cannot be applied to
frustrated spin systems. Actually, the estimated peak position of $17 J \approx 880$ cm$^{-1}$ with $s=3/2$ and $z=6$ is much higher than the high-energy cut-off of the magnetic spectral weight at around 440 cm$^{-1}$. The highly frustrated $s=1/2$ kagome lattice system ZnCu$_3$(OH)$_6$Cl$_2$ shows a broad continuum located at about $2~J$ with no clear cutoff but a moderate suppression at higher energy scales above $6~J$.~\cite{wulferding-10} This is also contrasted by the $s=1/2$ square lattice, which has a dominant spectral weight around $2.7~J$.

In a 2D triangular lattice, quantum corrections can lead to a substantial downscaling of the spectral weight due to the increased level of frustration and strongly modify the shape of the magnon dispersion.~\cite{perkins} For $s=1/2$, the one-magnon dispersion is characterized by shallow roton-like minima and a flat region extending over a wide range of the Brillouin zone. This leads to two van-Hove singularities in the Raman response. As a result, the bare two-magnon profile has two peaks. This feature will be smeared out due to magnon-magnon scatterings and especially the high energy peak is strongly suppressed due to larger magnon damping. Thus, a broad, rather symmetric Raman continuum appears with the round maximum around twice the roton energy.

For $s=3/2$, the spectral weight is expected to shift towards the higher energy region. In contrast, the two-magnon continuum of $\alpha$-CaCr$_2$O$_4$ gradually increases with decreasing energy. Its intensity is drastically suppressed for frequencies above 440 cm$^{-1}$ ($\approx~55$~meV) and the weak signal cannot be differentiated from the strong phonon peaks. The same features, in particular, the pronounced spectral weight at low energies, have been observed in the one-magnon spectrum by INS. They were attributed to mode softening due to the instability toward the neighboring new magnetic structure. Additionally, INS experiments show van-Hove singularities around 5, 11, 17, and 33 meV.
We mark twice these energies by black arrows on top of the two-magnon continuum (see the lower panel of Fig.~1). At the respective energies, we find the substantial spectral weights. To highlight the structured spectral weight of the magnetic scattering continuum, we plot the data together with three gaussian fits (black dashed lines) and a sum of these fits (blue, solid curve). The fits describe the data well. The three broad peaks identified in our data are related to low-energy roton-like modes. The minor discrepancies in energy are due to the different scattering matrix element between INS and Raman scattering. The multi-peak feature suggests that $\alpha$-CaCr$_2$O$_4$ is close to a new magnetic (multi-$k$) phase.

\begin{figure}
\label{figure3}
\centering
\includegraphics[width=8cm]{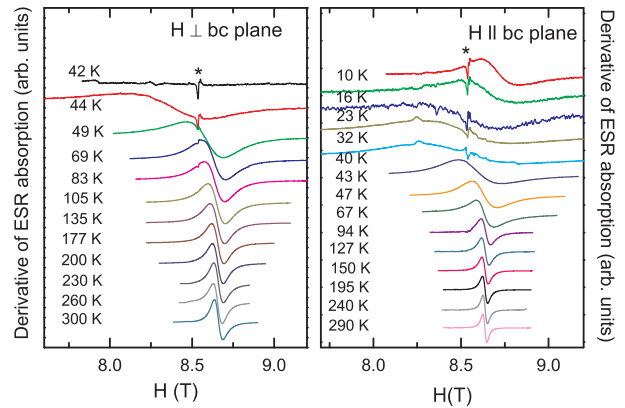}
\caption{(Colour online) Derivative of the ESR absorption of the $\alpha-$CaCr$_2$O$_4$ crystal measured at $\nu =240$ GHz for a field perpendicular (left panel) and parallel (right panel) to the $bc$ plane as a function of temperature. The spectra are vertically shifted for clarity. The asterisk denotes defects or orphan spins  since it exhibits no shift with temperature and no orientation dependence with a $g$-factor of $g=2$.}
\end{figure}

To obtain further information on the evolution of magnetic correlations we performed high-frequency ESR at $\nu =240$ GHz. Figure 3 displays the temperature dependence of ESR spectra for the field perpendicular and parallel to the $bc$ plane. At room temperature the ESR signal consists of a single lorentzian line, which originates from paramagnetic Cr$^{3+}$ ions. The exchange-narrowed resonance is usually expected for concentrated, insulating spin systems. We obtain the effective $g$-factors of $g_\perp=1.97(9)$ and $g_\|=1.98(6)$ at room temperature. The determined $g$-values are what is expected for a less than half-filled ion with a quenched orbital moment. Here, we note that the magnetic Raman scattering indicates that short-range correlated states persist well above room temperature (see the inset of Fig.~2). Thus, the intrinsic $g$ values in the paramagnetic limit are expected to increase slightly toward the free spin value. However, this difference will be marginal between room temperature and infinite temperatures. Therefore, the room temperature $g$-factors can be considered as asymptotic infinite-temperature values.

With decreasing temperature the spectrum undergoes a broadening and a shift. For $H \bot$ $bc$ plane, the signal wipes out for temperatures below $T_N$. In contrast, for $H \|$ $bc$ plane, we are able to keep track of the signal through $T_N$ down to the lowest measured temperature.
Noticeably, the weak, sharp peak (denoted by the asterisk) shows up for temperatures below 80~K. The $g$-factor of the extra peak is close to $g=2$ and independent of both orientation and temperature.
Furthermore, in an XRD analysis no obvious impurity phases were found~\cite{islam-mail} and Raman scattering spectra give no hint for lattice inhomogeneities. Thus, it is ascribed to a few percentages of defects and orphan spins, which are sensitive to ESR. Hereafter, we will concentrate on the main signal.
To detail the temperature dependence of the spin dynamics, the resonance field ($H_{res}$) and the peak-to-peak linewidth ($\Delta H_{pp}$) are extracted by fitting to a lorentzian profile.
The resulting out-of-plane (open triangle) and in-plane (full circle) components of $H_{res}$ and $\Delta H_{pp}$ are summarized in Fig.~4.

\begin{figure}
\label{figure4}
\centering
\includegraphics[width=8cm]{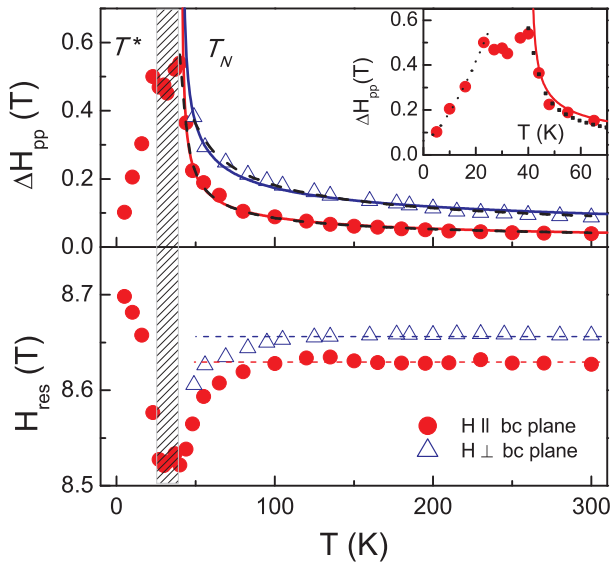}
\caption{(Colour online) Temperature dependence of the ESR linewidth $\Delta H_{pp}$ (upper panel) and the resonance field $H_{res}$ (lower panel).
The open triangle and the full circle symbols stand for $H_{res}$ and $\Delta H_{pp}$ for the orientation perpendicular and parallel to the $bc$ plane, respectively. Inset: A zoom of $\Delta H_{pp}$ at low temperatures. The solid lines are a fit to a power law and the dashed lines are a fit to a magnetic vortex model.}
\end{figure}

For $H \perp$ $bc$ plane, $H_{res}$ starts to shift to lower fields around 100~K. In a wide temperature region $\Delta H_{pp}$ exhibits a critical-like broadening. For $H \|$ $bc$ plane, $H_{res}$ starts to shift downward around 100~K and then shows a dip at about $T^{\ast}=24$~K, before finally shifting towards higher fields. Upon cooling down to $T_N$, $\Delta H_{pp}$ shows a critical broadening. At temperatures between $T_N$ and $T^{\ast}$, the linewidth exhibits saturation and below $T^{\ast}$, $\Delta H_{pp}$ drops but it has a residual value at the lowest measured temperature.

For an insulating, concentrated spin system the temperature dependence of the linewidth is given by $\Delta H_{pp}(\theta, T)=\alpha(\theta, T)\Delta H_{pp}(\theta,\infty)$ where $\Delta H_{pp}(\theta,\infty)$ is the linewidth in an uncorrelated paramagnetic limit. From the Kubo-Tomita formalism,~\cite{Kubo} we can estimate the linewidth of the ESR signal due to the spin-spin interactions in an uncorrelated paramagnetic regime as $$ \Delta H_{pp}(\theta,\infty)\approx \frac{1}{g\mu_B}\frac{M^2_2}{|J|},$$ where the second moment $M_2$ is a measure of the anisotropic contribution to the superexchange interaction $J=102$~K, assuming that the linewidth is dominated by the first order spin-orbital perturbation, $\Delta H_{pp}(\theta,\infty)\approx 4$~Oe. This leads to $\alpha(\theta, T)> 10$ at room temperature, implying that finite spin correlations persist up to room temperature. This is not surprising since 2D frustrated magnets exhibit the persistence of short-range spin correlations up to temperatures of the order of the Curie-Weiss temperature.~\cite{Yamaguchi08,Yamaguchi10} In our compound, the Curie-Weiss temperature is given by $\Theta_{CW}=-564$~K.~\cite{chapon, toth} This is further supported by the magnetic Raman scattering, which can still be observed at room temperature while retaining its spectral form.

To gain a deeper understanding of the spin relaxations, we analyze the temperature dependence of the linewidth using two models. The first model is based on a critical broadening given by $\Delta H_{pp}\propto (T-T_N)^{-p}$ with an exponent of $p \approx 0.385 (0.483)$ for the out-of-plane (in-plane) direction (see the solid lines in the upper panel of Fig.~4).
The observed exponent is half of the values of $p\approx 0.7-0.9$ reported in the triangular $s=3/2$ antiferromagnets $A$CrO$_2$ ($A$=Li,Na,Cu,Ag).~\cite{Hemmida09,Hemmida10,Hemmida11} The much smaller exponent indicates that the spin dynamics of $\alpha$-CaCr$_2$O$_4$ strongly deviates from what is expected for a triangular spin system with a 120$^\circ$ phase and is governed by higher-dimensional spin fluctuations.

The second model resorts to magnetic vortices as a spin relaxation channel. The 2D triangular Heisenberg antiferromagnet has been proposed to undergo a Kosterlitz-Thouless (KT) -like transition with Z$_2$ vortices.~\cite{Kawamura,Okubo,Misawa} The Z$_2$ vortices are related to the vector chirality defined by the 120$^\circ$ spin structure. Such a vortex scenario has been intensively examined in triangular chromium oxides that form a 120$^\circ$ ground state.~\cite{Hemmida09,Hemmida10,Hemmida11} Experimental indications of the KT phase transition have been obtained by studying the evolution of the ESR linewith in a paramagnetic phase.

The contribution of the vortex dynamics to the ESR linewidth is approximated by $\Delta H_{pp}\propto \xi(T)^3$, where $\xi(T)=\xi_0 \exp(b/\tau^{\nu})$ is the correlation length with $\tau=(T/T_{m}-1)$.~\cite{Mertens,Becker} The exponent $\nu$ depends on the specific model:
$\nu=0.5$ for the KT model and $\nu=0.37$ for the Kosterlitz-Thouless-Halperin-Nelson-Young model, which describes the 2D melting mechanism in the Coulomb gas on a triangular lattice.~\cite{Halperin,Young} The vortex model seems to give a satisfactory description of the linewidth in the whole paramagnetic range with the fit parameters $b=2.58(3.22)$, $T_{m}=28.9$ K (39.6 K), and $\nu=0.07 (0.05)$ (see the dashed lines of the upper panel of Fig.~4). The temperature $T_{m}$, which is related to the 2D melting transition, lies below $T_N$. Remarkably, the exponent $\nu$ is one order of magnitude smaller than the theoretical models suggest. Since the Cr$^{3+}$ ions occupy half-filled $t_{2g}$ orbitals, the anisotropies will be very small. Indeed, the $g$-factor anisotropy is about 1-2\%, confirming the minute exchange and single ion anisotropies. This means that $\alpha$-CaCr$_2$O$_4$ is of dominant Heisenberg character. Therefore, in our case the anisotropies cannot explain the extremely small exponent.

We recall that the Cr-based triangular antiferromagnets $A$CrO$_2$ ($A$=H,Li,Na,Cu,Ag), which have a comparable size of anisotropies to $\alpha$-CaCr$_2$O$_4$ are well described by the vortex model with $\nu\approx 0.36-0.47$.~\cite{Hemmida09,Hemmida10,Hemmida11} The much smaller exponent suggests that the vortex dynamics is irrelevant to our system although the vortex model yields a satisfactory fit to the linewidth. In contrast to $A$CrO$_2$, in our system the Z$_2$ vortices will not be formed in spite of a 120$^\circ$ spin structure. This further supports the finding of the instability of a 120$^\circ$ phase due to longer-range exchange interactions.

Regarding the temperature dependence of the $g$-factors (the resonance field), our system shows a downshift of the resonance field at temperatures well above $T_N$. Since we rule out the vortex scenario, this shift is related to the formation of internal fields due to short-ranged magnetic ordering, which is a characteristic feature of frustrated magnets.~\cite{Viticoli}

Next, we will turn to an antiferromagnetic resonance (AFMR) mode which appears for temperatures below $T_N$. For conventional antiferromagnets AFMR arises from a spin wave excitation by a microwave. The temperature dependence of the AFMR linewidth is determined by the population of magnons and is phenomenologically described by a power law $\Delta H_{pp}\propto T^{4}$.~\cite{Rezende}
In case of our system, the linewidth increases continuously down to $T_N$ and exhibits saturation between $T^{\ast}$ and $T_N$.
For temperatures below $T^{\ast}$ the drop of the AFMR linewidth is described by $\Delta H_{pp}\propto T^{1.48}+\Delta H_{pp}(0)$ with a residual linewidth of $\Delta H_{pp}(0)=0.064$~T (see the inset of the upper panel of Fig.~4). The constant linewidth between $T^{\ast}$ and $T_N$ implies a gradual freezing into a long-range ordered state, which is a characteristic feature of frustrated magnets.~\cite{Olariu} The weaker temperature dependence of the AFMR linewidth indicates the presence of residual quantum fluctuations and enhanced magnon-magnon interactions for a triangular lattice. This is corroborated by the monotonic increase of the two-magnon intensity for temperatures below $T_N$ as shown in the inset of Fig. 2.

Lastly, we will stress the peculiar spin topology derived from unusual distortions of a triangular lattice. The four inequivalent nearest neighbor interactions form two zig-zag and two linear chains. The set of magnetic parameters are still in the range to stabilize a 120$^\circ$ structure. However, low-energy magnetic excitations will be modified due to next-nearest neighbor interactions and competing spin networks.
The partially lifted frustration leads to a softening of the van-Hove singularities.
Contrary to other Cr-based $s=3/2$ triangular compounds, the Z$_2$ vortices and chiral fluctuations will not be well-defined. This provides a natural explanation for the softened two-magnon continuum and the strongly reduced critical exponent of the ESR line broadening.

\section{Summary}
In summary, we have presented a combined Raman scattering and ESR study of the $s=3/2$ distorted triangular antiferromagnet $\alpha$-CaCr$_2$O$_4$. In spite of the helical 120$^\circ$ spin structure, this system shows distinct anomalies in magnetic excitations and spin dynamics. The roton dynamics lead to a two-magnon continuum with a multi-maximum structure and a softening of the spectral weight. The critical ESR line broadening is described by a comparably small exponent. This is ascribed to an instability of this system due to its proximity to a new magnetic phase, which is caused by the complex distortions from a perfect triangular lattice and the presence of substantial next-nearest neighbor interactions.

\begin{acknowledgments}
We acknowledge fruitful discussions with W. Brenig and support from DFG, B-IGSM and the NTH School Contacts in Nanosystems. KYC acknowledges financial support from Humboldt foundation and Korea NRF Grant No. 2009-0093817.
\end{acknowledgments}

\end{document}